\documentclass[letterpaper]{jpsj-suppl}
\usepackage{txfonts} %Please comment out this line unless the txfonts package is availabe in your LaTeX system.
\usepackage{multirow}
\usepackage{capt-of}
\usepackage{cite}
\def\newblock{\hskip .11em plus .33em minus .07em}

\title{An Intermediate Water Cherenkov Detector at J-PARC}

\author{Mark \textsc{Scott}$^{1}$ for the NuPRISM and Hyper-K collaborations}

\inst{$^{1}$TRIUMF, Vancouver, British Columbia, Canada}

\email{mscott@triumf.ca}

\recdate{January 31, 2016}

\abst{Recent neutrino oscillation results have shown that the existing long baseline experiments have some sensitivity to the effects of CP violation in the neutrino sector.  This sensitivity is currently statistically limited, but the next generation of experiments, DUNE and Hyper-K, will provide an order of magnitude more events.  To reach the full potential of these datasets we must achieve a commensurate improvement in our understanding of the systematic uncertainties that beset them.
 
This talk describes two proposed intermediate detectors for the current and future long baseline oscillation experiments in Japan, TITUS and NuPRISM.  These detectors are discussed in the context of the current T2K oscillation analysis, highlighting the ways in which they could reduce the systematic uncertainty on this measurement.  The talk also describes the short baseline oscillation sensitivity of NuPRISM along with the neutrino scattering measurements the detector makes possible.}

\begin{document}
\maketitle

\section{Introduction} 
\label{intro}

Long baseline neutrino oscillation experiments have reached the point where systematic uncertainties dominate over statistical precision when measuring neutrino disappearance.  The discovery of large $\theta_{\textrm{13}}$ by Daya Bay~\cite{PhysRevLett.112.061801}, RENO~\cite{Ahn:2012nd} and T2K~\cite{PhysRevLett.112.061802}, combined with increases in neutrino beam power and the construction of larger experiments, mean that this will soon be the case for neutrino appearance measurements as well.  The sensitivity of future experiments to CP violation will depend strongly on how well they can control their systematics.

\begin{figure}[ht!]
\begin{minipage}[t]{.46\textwidth}
\includegraphics[height=5cm]{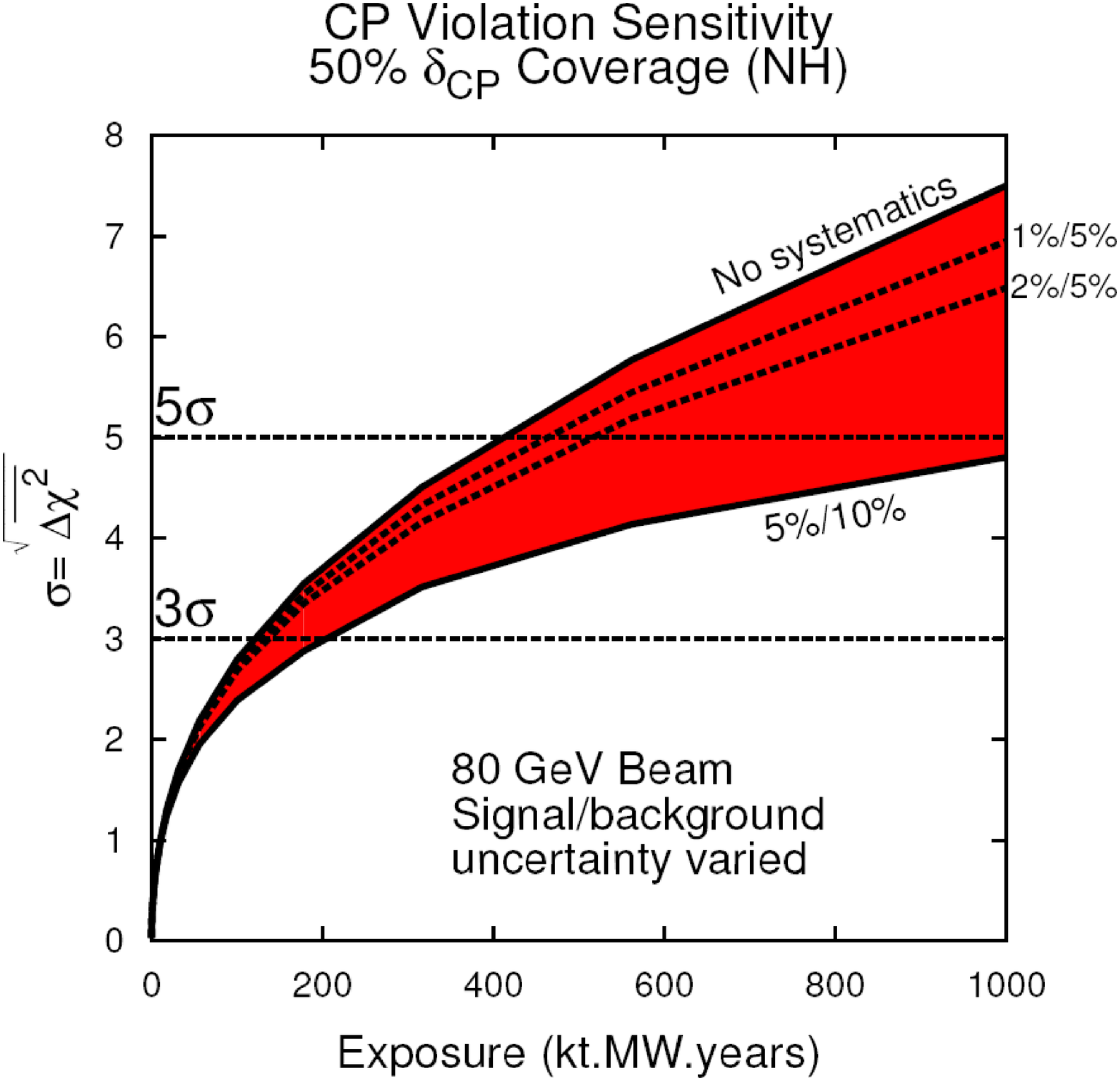}
\centering\\
Predicted sensitivity to CP violation of the LBNE experiment over 50\% of the $\delta_{\scriptscriptstyle\rm{CP}}$ parameter space, assuming a range of signal and background normalisation uncertainties, as a function of exposure.~\cite{Adams:2013qkq}
\end{minipage} \hfill
\begin{minipage}[t]{.46\textwidth}
\includegraphics[height=4.5cm]{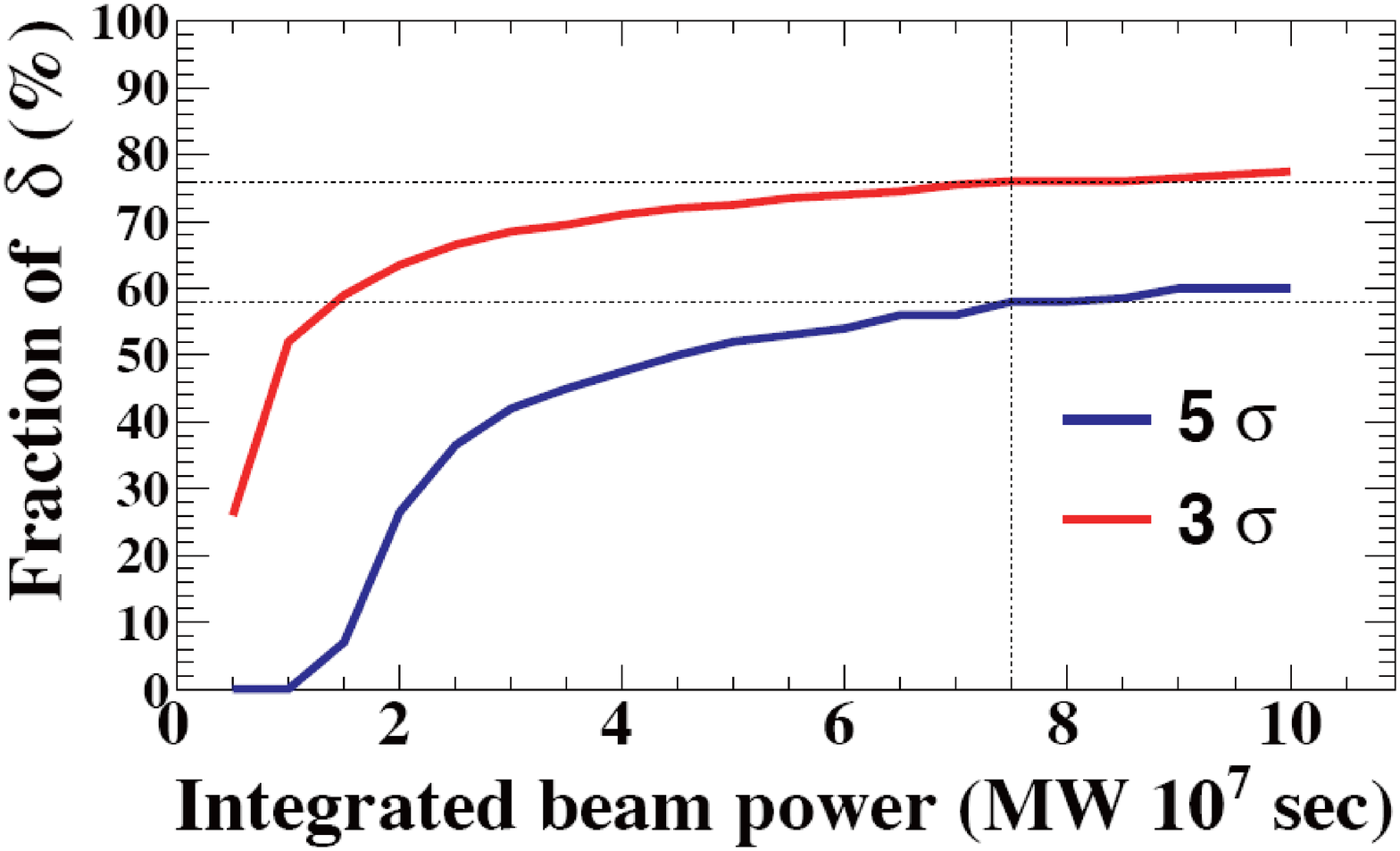}
\centering\\
Predicted fraction of $\delta_{\scriptscriptstyle\rm{CP}}$ parameter space for which a 3$\sigma$~(red) or 5$\sigma$~(blue) measurement of CP violation would be made for a given integrated beam power at Hyper-K.  This assumes a total uncertainty of 3\% on the far detector event rate.~\cite{Abe:2015zbg}
\end{minipage}
\caption{Published CP violation sensitivity curves from the LBNE (left) and Hyper-K (right) experiments.}
\label{fig:sensitivities}
\end{figure}

The next generation experiments, DUNE~\cite{Acciarri:2015uup} and Hyper-Kamiokande (Hyper-K)~\cite{Abe:2014oxa}, require the total systematic uncertainty on their far detector rate prediction to be less than 3\%, as shown by Figure~\ref{fig:sensitivities}.  This can be compared to the systematic uncertainty currently demonstrated by the T2K experiment~\cite{PhysRevD.91.072010}, which has achieved a systematic uncertainty of around 7\% on their far detector event rate prediction, with the greatest part of this coming from nuclear interaction uncertainties.

The T2K analysis parameterises both their neutrino flux prediction and their neutrino interaction model, producing a set of parameters with associated errors that are then constrained using data from the T2K near detector, ND280.  This produces a tuned prediction for the far detector event rate, changing the central value of the flux and cross section model parameters whilst reducing their uncertainty.  The far detector event rate uncertainties are shown in Table~\ref{tab:T2K_Syst}.

\begin{table}
\centering
\begin{tabular}{@{}llcc@{}}
\hline
         Source of uncertainty                           &                                         & $\nu_{\mu}$ sample & $\nu_{e}$ sample \\ \hline
\multirow{2}{*}{Flux and common cross section}           & w/o ND measurement                      & 21.7\%             & 26.0\%           \\
                                                         & w/ ND measurement                       & 2.7\%              & 3.2\%            \\ \hline
\multicolumn{2}{p{7cm}}{Independent cross sections}                                                & 5.0\%              & 4.7\%            \\ 
\multicolumn{2}{p{7cm}}{Super-K detector}                                                          & 4.0\%              & 2.7\%            \\ 
\multicolumn{2}{p{7cm}}{Final or Secondary Hadronic Interaction}                                   & 3.0\%              & 2.5\%            \\ \hline
\multirow{2}{*}{Total}                                   & w/o ND measurement                      & 23.5\%             & 26.8\%           \\
                                                         & w/ ND measurement                       & 7.7\%              & 6.8\%            \\ \hline
\end{tabular}
\caption[margin=5cm]{Table showing the uncertainty on the predicted number of selected events at the T2K far detector, broken down by source~\cite{PhysRevD.91.072010}.}
\label{tab:T2K_Syst}
\end{table}

Firstly, Table~\ref{tab:T2K_Syst} shows that near detectors are essential to reduce the effect of flux and neutrino interaction cross-section systematics at the far detector, with the far detector event rate uncertainty falling from 24\% to 3\% because of the near detector constraint.  Table~\ref{tab:T2K_Syst} also shows that the largest far detector uncertainty is caused by `Independent cross section' systematics.  These are associated to neutrino interaction processes that, for two main reasons, the T2K near detector did not measure in this analysis:
\begin{enumerate}
    \item{Different target nuclei at the near and far detectors}
    \item{A near detector insensitive to some far detector backgrounds}
\end{enumerate}
Few of the recent neutrino cross-section measurements have been made with an oxygen target, and there are significant uncertainties on the scaling of the cross section between different nuclei.  The T2K near detector has two targets, one fully composed of plastic scintillator and the second a combination of plastic scintillator and water.  The analysis discussed above used data from the plastic scintillator target, so could not constrain the interaction cross section on oxygen.  Future T2K analyses will also include data from the water target, fitting both carbon and oxygen interactions simultaneously.  This will provide a constraint on the neutrino interaction cross section on oxygen, but will be fundamentally limited by the need to statistically subtract interactions on carbon from the water sample.

For the second point, without samples of the far detector background processes it is impossible to constrain their associated uncertainties using near detector data.  Using the same detection technology at both near and far detectors would ensure that the background events in the far detector can be measured at the near detector.

In addition to the points above, the T2K near detector geometry was optimised to reconstruct particles travelling in the same direction as the incoming neutrino beam.  As a consequence it has almost no acceptance for particles travelling perpendicularly to the neutrino beam.  Meanwhile the far detector, Super-Kamiokande (SK), has a high reconstruction efficiency across the full 4$\pi$ solid angle.  T2K must therefore rely on their neutrino interaction model to extrapolate the reduced phase space near detector data to the full phase space observed by SK.  This will limit how far the systematics in the T2K oscillation analysis can be reduced.

Luckily, these limitations can be overcome in the future by building a water Cherenkov detector between 1~km and 2~km from the T2K neutrino beam production point.

\section{An intermediate water Cherenkov detector}

Building a kiloton scale water Cherenkov detector around 1~km or 2~km from the T2K neutrino beam production point provides four benefits:
\begin{enumerate}
    \item{A water target}
    \item{Identical signal and background interaction modes as at SK}
    \item{4$\pi$ solid angle acceptance}
    \item{A smaller error on the flux extrapolation from the near to the far detector}
\end{enumerate}

An identical target and detection technology mean that this intermediate detector will address the the issues discussed in Section~\ref{intro}.  The T2K neutrino beam is created by pion decay-in-flight, with pions produced by impinging protons from the J-PARC main ring onto a carbon target.  The pions are then focussed into a 90~m long volume, where they decay to produce neutrinos.  The ND280 detector is 280~m downstream of the carbon target, so measures a line source of neutrinos.  SK, 295~km away, observes the neutrinos as if from a point source. This difference in the neurino spectrum at the near and far detectors means that the near-to-far flux extrapolation is imperfect.  By siting an intermediate detector further from the neutrino production point it will see a flux much more similar to that at SK than at the ND280, reducing the uncertainty in the flux extrapolation.

This talk discusses two proposed intermediate detectors for the J-PARC neutrino program, TITUS~\cite{Lasorak:2015eba} and NuPRISM~\cite{Bhadra:2014oma}.

\subsection{TITUS}

\begin{figure}[ht!]
\centering
\begin{minipage}[t]{.6\textwidth}
\includegraphics[height=5cm]{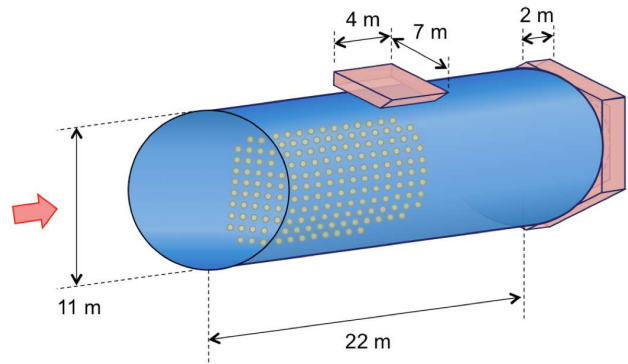}
\caption[margin=5cm]{An initial design of the proposed TITUS detector.}
\label{fig:titus}
\end{minipage}
\end{figure}

TITUS, the \textbf{T}okai \textbf{I}ntermediate \textbf{T}ank to measure the \textbf{U}noscillated \textbf{S}pectrum, is a cylindrical water Cherenkov detector with its long axis parallel to the neutrino beam, shown in Figure~\ref{fig:titus}.

The detector would be placed 2~km from the neutrino production point and be instrumented with with PMTs interspersed with large area picosecond photo-detectors if these become available.  The design also features two magnetised muon range detectors (MRDs), one downstream of the tank and the other on the top edge.  TITUS was designed to perform neutron tagging, so incorporates a 0.1\% by mass gadolinium loading in the baseline design.

The TITUS studies presented here take the particle reconstruction efficiency and resolution from the SK detector simulation.  The detector response model is calculated as a function of distance of the most energetic particle to the wall, taking into account the smaller size of TITUS relative to SK.  This process assumes that the TITUS reconstruction will be able to achieve the same performance as SK.  The existing SK 1-ring electron-like and 1-ring muon-like selections~\cite{PhysRevD.91.072010} are then applied to give the selected TITUS samples.

\subsubsection{Detector orientation}

\begin{figure}[ht!]
\centering
\includegraphics[height=5cm]{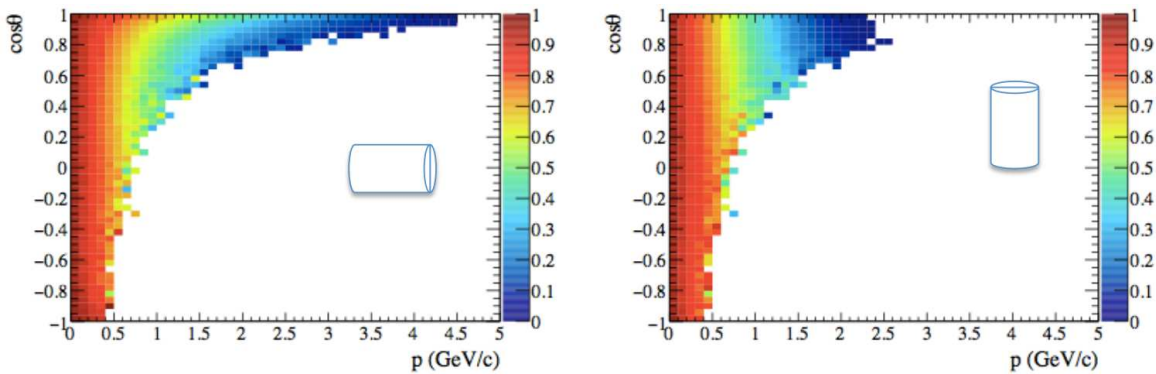}
\caption[margin=5cm]{The muon reconstruction efficiency as a function of muon momentum and angle to the neutrino beam axis for the two proposed detector orientations.  The muon is required to stop within the detector for it to be reconstructed.}
\label{fig:titus_acceptance}
\end{figure}

The TITUS group studied the muon reconstruction efficiency for two detector orientations -- one with the long axis parallel to the neutrino beam the other with the axis perpendicular to the beam.  Figure~\ref{fig:titus_acceptance} shows the efficiency for these two situations as a function of the muon momentum and angle to the neutrino beam axis. 

The vertical tank is unable to reconstruct muons with momenta greater than 2.5~GeV/c, and loses some efficiency as the muon direction approaches the radial direction of the cylinder.  This study motivated the choice for the detector orientation, but even so 18\% of the muons from charged current neutrino interactions are expected to exit the detector.  These exiting muons can be recovered by the muon range detectors.

\subsubsection{Magnetised Muon Range Detectors}

The MRDs are tracking detectors composed of iron sheets interleaved with scintillator layers and air gaps, magnetised to 1.5~T.  By tuning the iron thickness and the size of the air gaps the MRDs will be able to use the curvature of particles to measure their charge, achieving a 90--95\% efficiency for muons with momenta from 0.5--2~GeV/c.  Placing an MRD at the downstream end of the water tank will allow higher momentum particles, that would usually exit the detector,  to be reconstructed correctly.  Similarly, a smaller MRD on the side of the tank provides additional acceptance for muons travelling perpendicularly to the neutrino direction.

A proof-of-principal detector is being constructed by the University of Geneva~\cite{Asfandiyarov:2014haa} for use in the WAGASCI experiment~\cite{Koga:2015iqa} at J-PARC.  This data will then be used to optimise the MRD design for TITUS.

\subsubsection{Gadolinium doping}

\begin{figure}[ht!]
\centering
\begin{minipage}[t]{.46\textwidth}
\includegraphics[height=5cm]{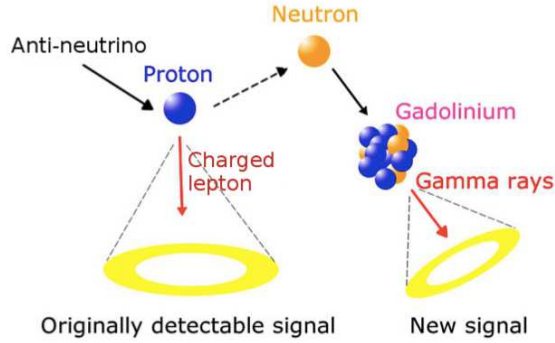}
\caption[margin=5cm]{A cartoon showing an anti-neutrino interaction followed by the capture of the produced neutron on gadolinium.}
\label{fig:gd_cartoon}
\end{minipage}
\end{figure}

Gadolinium has a neutron capture cross section of 49,000~barns, far greater than for neutron capture on hydrogen.  The capture process produces an excited state of gadolinium which promptly decays by emitting an 8~MeV gamma cascade of which 4-5~MeV is visible in a water Cherenkov detector.  A cartoon of an anti-neutrino interacting with a proton is shown in Figure~\ref{fig:gd_cartoon} to illustrate this process.
The SK collaboration has recently decided to introduce gadolinium to the SK detector, so in order to have the same target composition both TITUS and NuPRISM have included gadolinium doping in their design.

\begin{figure}[ht!]
\centering
\includegraphics[height=5cm]{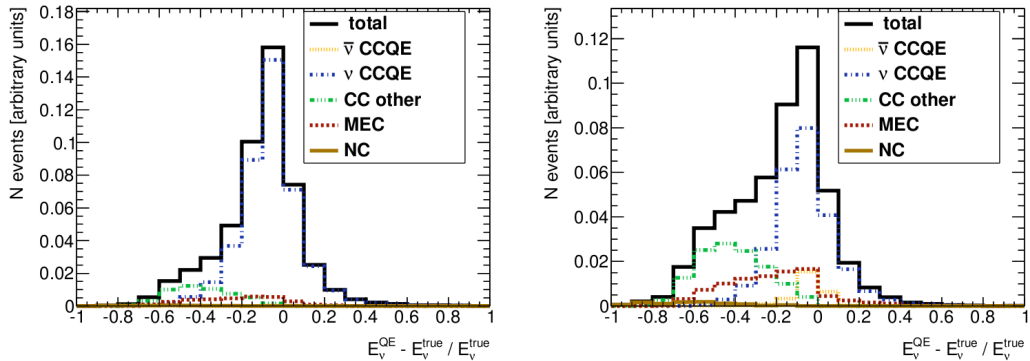}
\caption[margin=5cm]{The reconstructed neutrino energy resolution, assuming charged current quasi-elastic kinematics, for events selected at TITUS from the T2K neutrino mode beam.  The selected events are separated according to the number of tagged neutrons, 0 on the left and $>$ 1 on the right.}
\label{fig:gd_study}
\end{figure}

For neutrino oscillation studies, gadolinium enables neutrino interactions to be categorised by the number of neutrons in the final state.  
This could allow neutrino charged current quasi-elastic (CCQE) interactions to be statistically separated from other processes by selecting events with 0 tagged neutrons.  
Initial studies of this have been done with TITUS and are shown in Figure~\ref{fig:gd_study}.

Given the neutron multiplicities predicted by the neutrino event generator used to simulate these interactions (NEUT~\cite{Hayato:2009zz}), the combination of gadolinium loading and the MRDs allows TITUS to select samples of neutrino and anti-neutrino charged current events with a purity of 96\% at the T2K neutrino flux peak.  More usefully, the MRDs provide a check of the neutron tag by correlating the number of observed neutrons with the charge of the observed lepton.  Such a validation would be much less dependent on the correct simulation of poorly understood nuclear effects.

\subsubsection{Hyper-K CPV sensitivity study with TITUS}

\begin{figure}[!h]
\begin{minipage}[t]{.52\textwidth}
\vspace{50pt}
\begin{center}
\begin{tabular}{ l l }
    \hline
    Parameter & Nominal value and prior uncertainty \\ \hline
    $\delta_{\scriptscriptstyle\rm{CP}}$ & $0.000$, uniform in $\delta_{\scriptscriptstyle\rm{CP}}$ \\
    sin$^{2}2\theta_{13}$ & $0.095$, uniform in sin$^{2}2\theta_{13}$ \\
    sin$^{2}2\theta_{23}$ & $1.000 \pm 0.03$, \\
    sin$^{2}2\theta_{12}$ & $0.857 \pm 0.034$ \\
    $\Delta \textrm{m}^{2}_{32}$ & $2.320 \pm 0.100 \times 10^{-3}$ eV$^{2}$ \\
    $\Delta \textrm{m}^{2}_{12}$ & $7.500 \pm 0.200 \times 10^{-5}$ eV$^{2}$ \\
    \hline
\end{tabular}
\captionof{table}[margin=5cm]{Oscillation parameters used in the TITUS $\delta_{\scriptscriptstyle\rm{CP}}$ sensitivity study.}
\label{tab:titus_osc}
\end{center}
\end{minipage} \hfill
\centering
\begin{minipage}[t]{.46\textwidth}
\vspace{0pt}
\includegraphics[height=5cm]{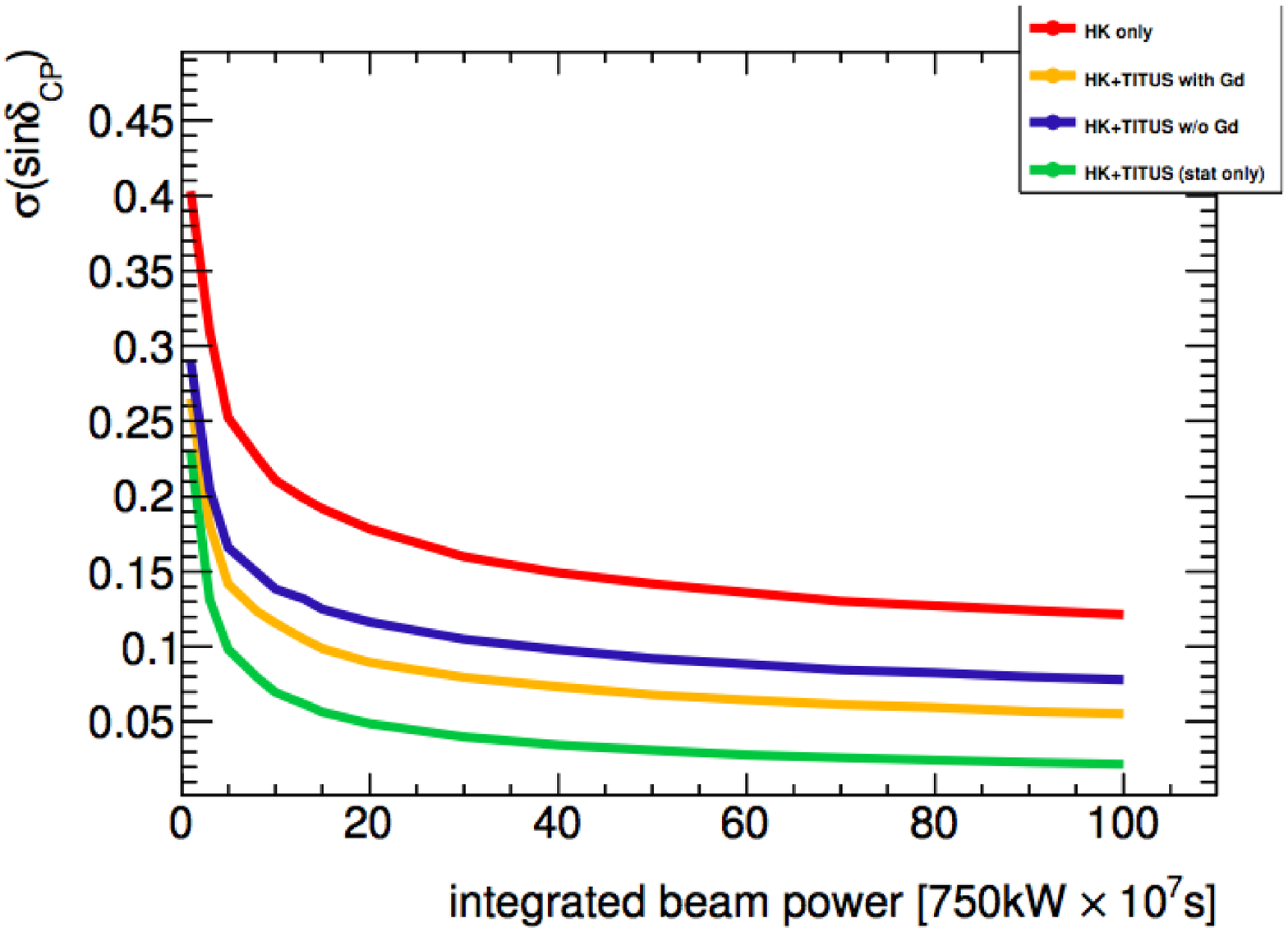}
\caption[margin=5cm]{The $1~\sigma$ uncertainty on the measured value of $\textrm{sin}~\delta_{\scriptscriptstyle\rm{CP}}$ as a function of integrated beam power for combinations of the Hyper-K and TITUS detectors with and without using neutron tagging as a selection cut.}
\label{fig:titus_cpv}
\end{minipage}
\end{figure}

A simplified oscillation analysis was performed to assess the impact of the TITUS detector on the Hyper-K $\delta_{\scriptscriptstyle\rm{CP}}$ sensitivity.  This was performed by simultaneously fitting samples of single ring, muon-like and electron-like events from the T2K neutrino and anti-neutrino beams, giving four samples in total at both TITUS and the Hyper-K far detector -- the T2K near detector is not included.  The study assumes a 6\% flux uncertainty that is 100\% correlated between TITUS and Hyper-K and 60\% correlated between the neutrino and anti-neutrino beam modes.  The T2K neutrino interaction model uncertainties~\cite{PhysRevLett.111.211803} were used and a 10\% uncertainty was assumed for the gadolinium neutron tagging efficiency.  The oscillation parameter values used in the study are shown in Table~\ref{tab:titus_osc}.  It is worth pointing out that this is the best case scenario, since the effect of gadolinium on the event reconstruction has not been included and the NEUT final state nucleon predictions are assumed to be correct.

The precision of measuring $\delta_{\scriptscriptstyle\rm{CP}} = 0$ for this setup is displayed in Figure~\ref{fig:titus_cpv}, which shows the fitted uncertainty on sin~$\delta_{\scriptscriptstyle\rm{CP}}$ as a function of the integrated neutrino beam power.  The integrated beam power is assumed to have been divided equally between the neutrino and anti-neutrino beam modes.

Figure~\ref{fig:titus_cpv} shows that adding the binary neutron tag discussed above to both the near and far detectors can lead to a 17\% improvement in the precision of a sin~$\delta_{\scriptscriptstyle\rm{CP}}$ measurement at Hyper-K.  This motivates a more detailed study into the hadronic side of neutrino interaction models, leading to better theoretical predictions for the hadronic final states and improved descriptions of particle re-interaction with the target nucleus.  Improved theoretical understanding must also be matched by improved experimental measurements of these processes in order to be confident enough to use these neutron tagging techniques in oscillation analyses.

\subsection{NuPRISM}

\begin{figure}
\begin{minipage}{.46\textwidth}
\includegraphics[height=5cm]{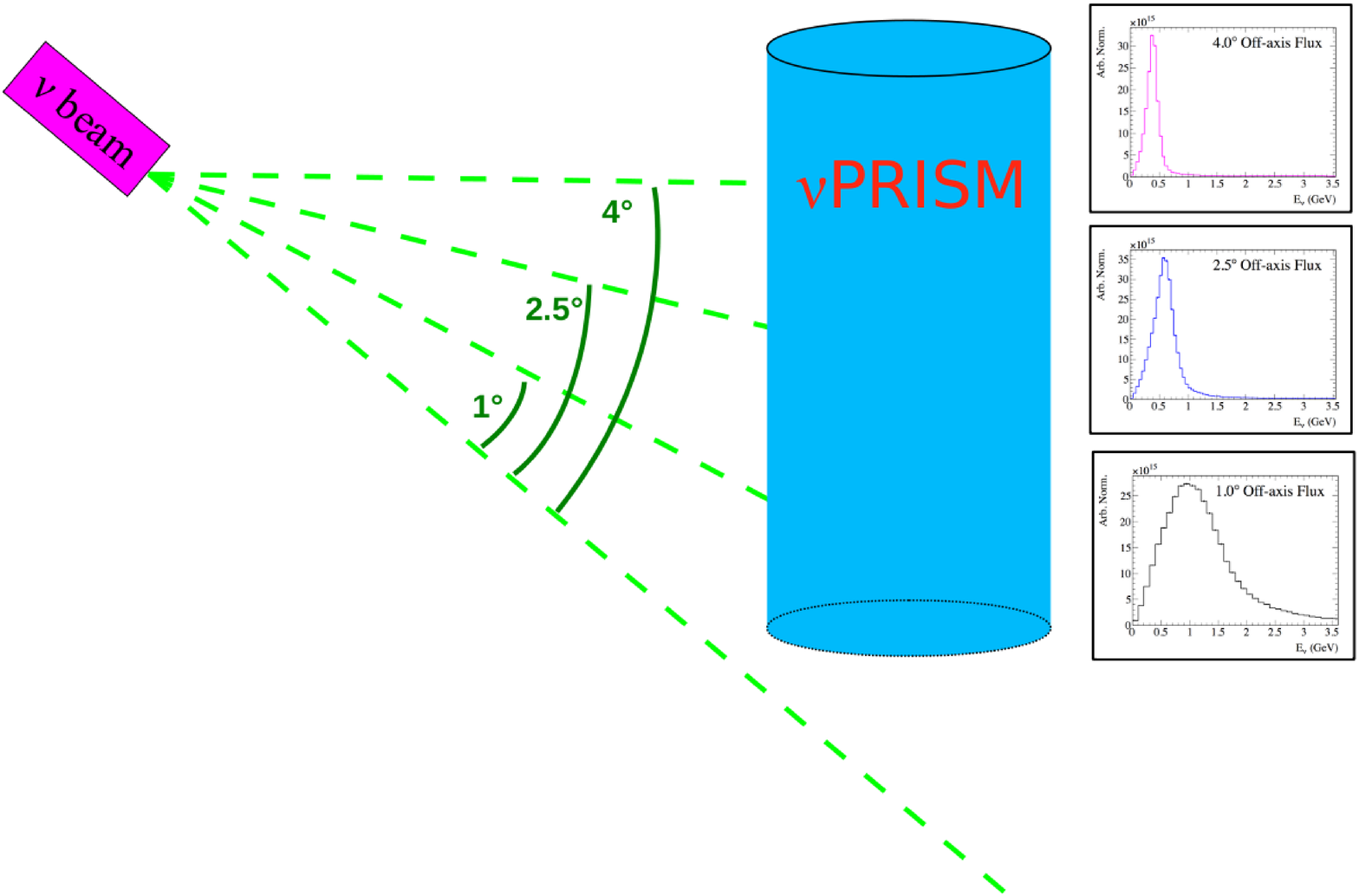}
\centering
\caption[margin=5cm]{The different neutrino energy spectra across the NuPRISM detector.}
\label{fig:flux_var}
\end{minipage}\hfill
\begin{minipage}{.46\textwidth}
\centering
\includegraphics[height=5cm]{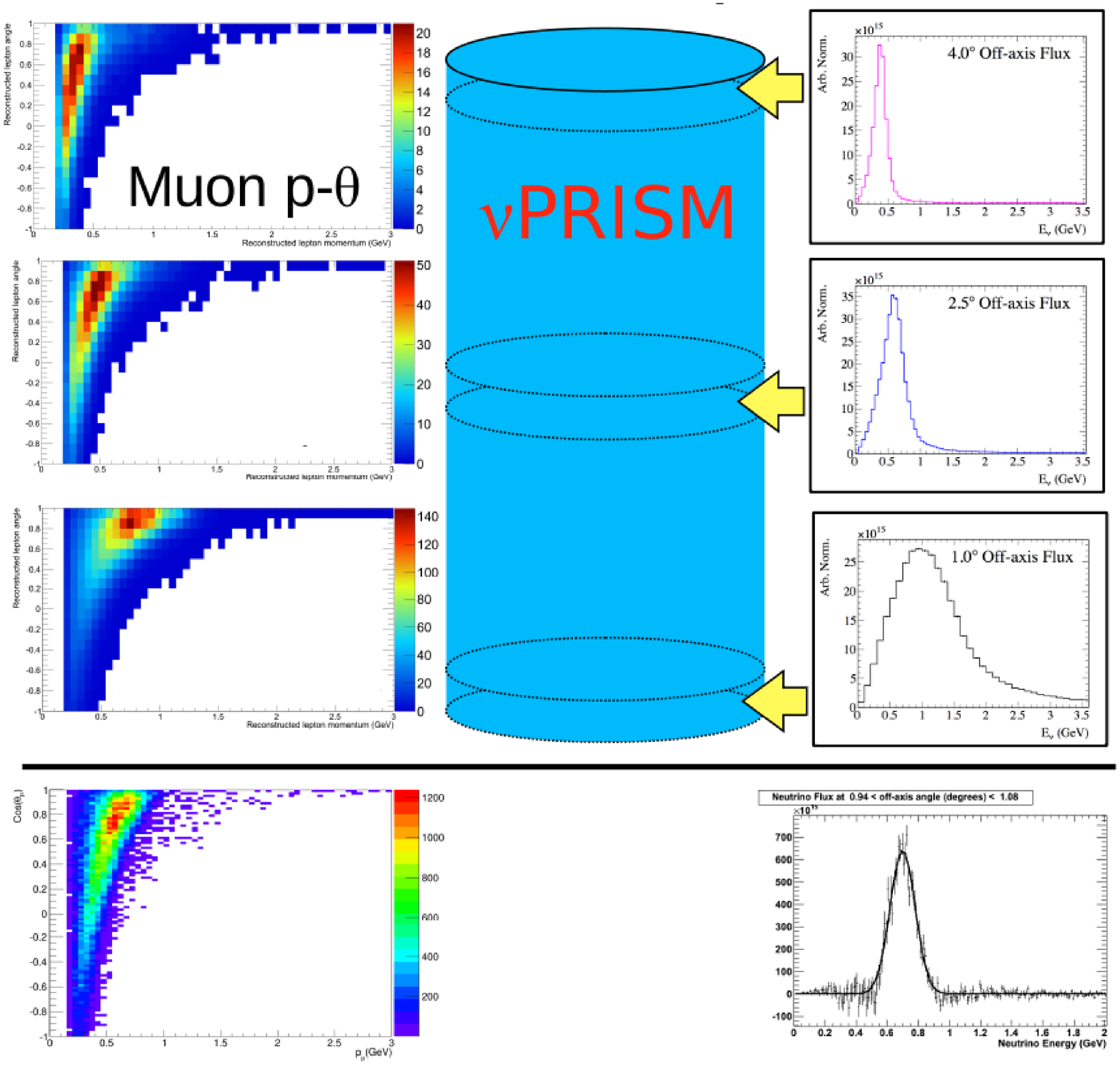}
\caption[margin=5cm]{An example of the linear combinations required to produce a Gaussian neutrino flux.}
\label{fig:gaus_flux}
\end{minipage}
\end{figure}

The neutrinos in a conventional neutrino beam come from the two-body decay-in-flight of charged pions.  As one moves further from the beam axis the observed neutrino energy spectrum narrows and peaks at a lower energy; this is called the ``off-axis'' effect.  By measuring neutrino interactions across a range of off-axis angles NuPRISM would sample many different neutrino spectra, each of which peaks at a different energy.  A cartoon of this is shown in Figure~\ref{fig:flux_var}.  The detector is split into slices, each at a different off-axis angle, which can be weighted and combined to create an arbitrarily shaped neutrino spectrum.  Reconstructed events are selected in each slice, and applying the chosen linear combination to these events gives the expected reconstructed event distribution for the desired neutrino flux.  An example of this is shown in Figure~\ref{fig:gaus_flux}, where a Gaussian flux centred at 700~MeV is created.  The 1D histograms on the right show the different off-axis fluxes whilst the 2D histograms show the corresponding reconstructed lepton momentum and angle to the neutrino beam.  The two lowest plots show the result of applying the linear combination, with the Gaussian flux on the right and the expected lepton kinematic distribution for that flux on the right.  Using this technique, NuPRISM provides a direct link between the observed reconstructed event information and the neutrino energy.

In this talk MC analyses were performed using a dataset corresponding to an exposure of $4.5e^{20}$ protons-on-target for each off-axis slice of NuPRISM.  This is equal to roughly half the expected T2K neutrino beam mode dataset and only 20\% of that proposed for the T2K-II extension.  All analyses presented here use the SK reconstruction efficiency to create the NuPRISM samples.  The efficiency at SK was calculated as a function of true lepton momentum, true angle and the distance from the interaction vertex to the closest wall of the SK tank.  The default NuPRISM design has a 3~m inner detector radius, so the reconstruction efficiency is taken from the outer 3~m ring of SK, where its performance is worst.  The NuPRISM group is working on a full detector simulation and reconstruction, so expect the reconstruction performance to improve for future analyses.
Individual analyses also incorporate the T2K flux and neutrino interaction uncertainties when needed, using the models from Ref.~\cite{PhysRevLett.111.211803}.  

\subsubsection{Gaussian neutrino beams}

\begin{figure}
\begin{minipage}{.46\textwidth}
\centering
\includegraphics[height=5cm]{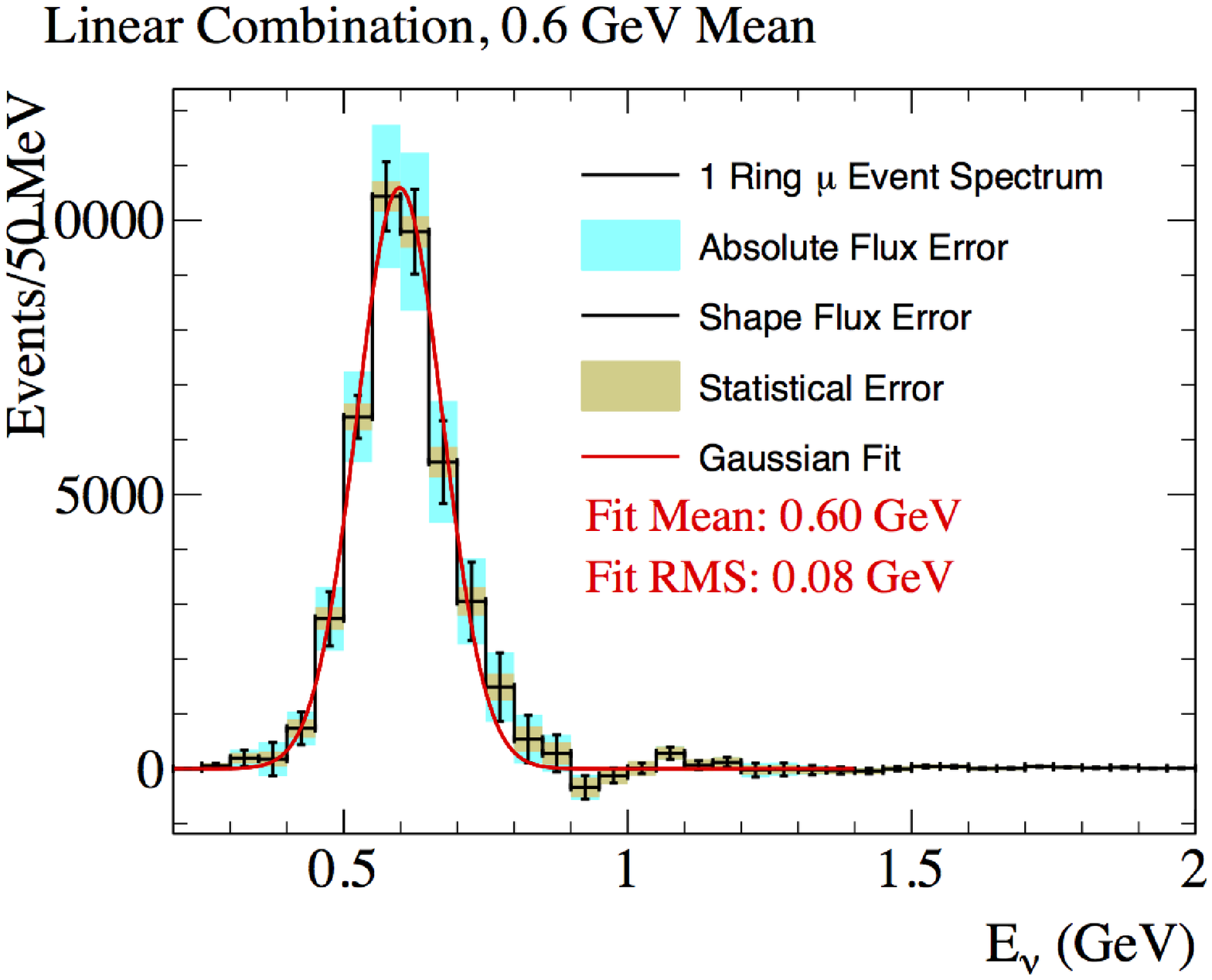}\\
a) 600~MeV.
\end{minipage}\hfill
\begin{minipage}{.46\textwidth}
\centering
\includegraphics[height=5cm]{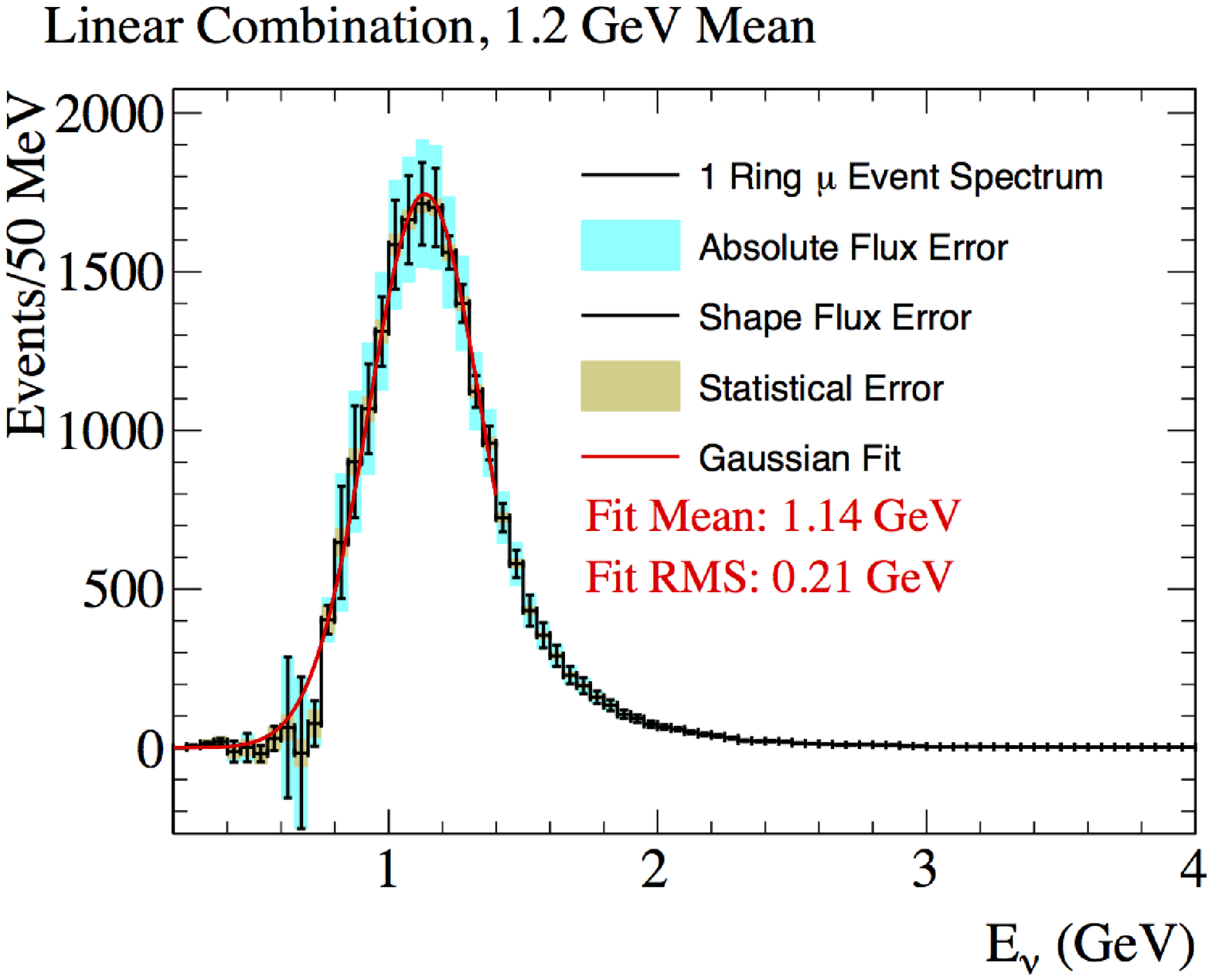}\\
b) 1200~MeV.
\end{minipage}
\caption[margin=5cm]{The true energy of the neutrinos that pass the NuPRISM single ring, muon-like selection after applying the linear combinations necessary to produce a Gaussian flux peaked at either 600~MeV (left) or 1200~MeV (right).  The flux systematic error is shown, along with the statistical uncertainty.}
\label{fig:gaus_enu}
\end{figure}

The range of neutrino energies for which NuPRISM can form a Gaussian flux is determined by the off-axis angles that NuPRISM spans.  The initial design has NuPRISM covering the $1$~--~$4~^\circ$ off-axis angles, allowing the creation of Gaussian beams from 400~MeV to 1200~MeV.  Figure~\ref{fig:gaus_enu} shows the true neutrino energy distribution of selected events for Gaussian neutrino fluxes centred at 600~MeV and 1200~MeV.  The light blue error bars depict the error on the absolute flux prediction, which is fully correlated across all bins, while the black error bars give the uncertainty on the flux shape.  The statistical uncertainty on the NuPRISM sample is shown by the light brown shading.  More details of this analysis can be found in Ref.~\cite{Bhadra:2014oma}.

\begin{figure}
\begin{minipage}{.46\textwidth}
\includegraphics[height=5cm]{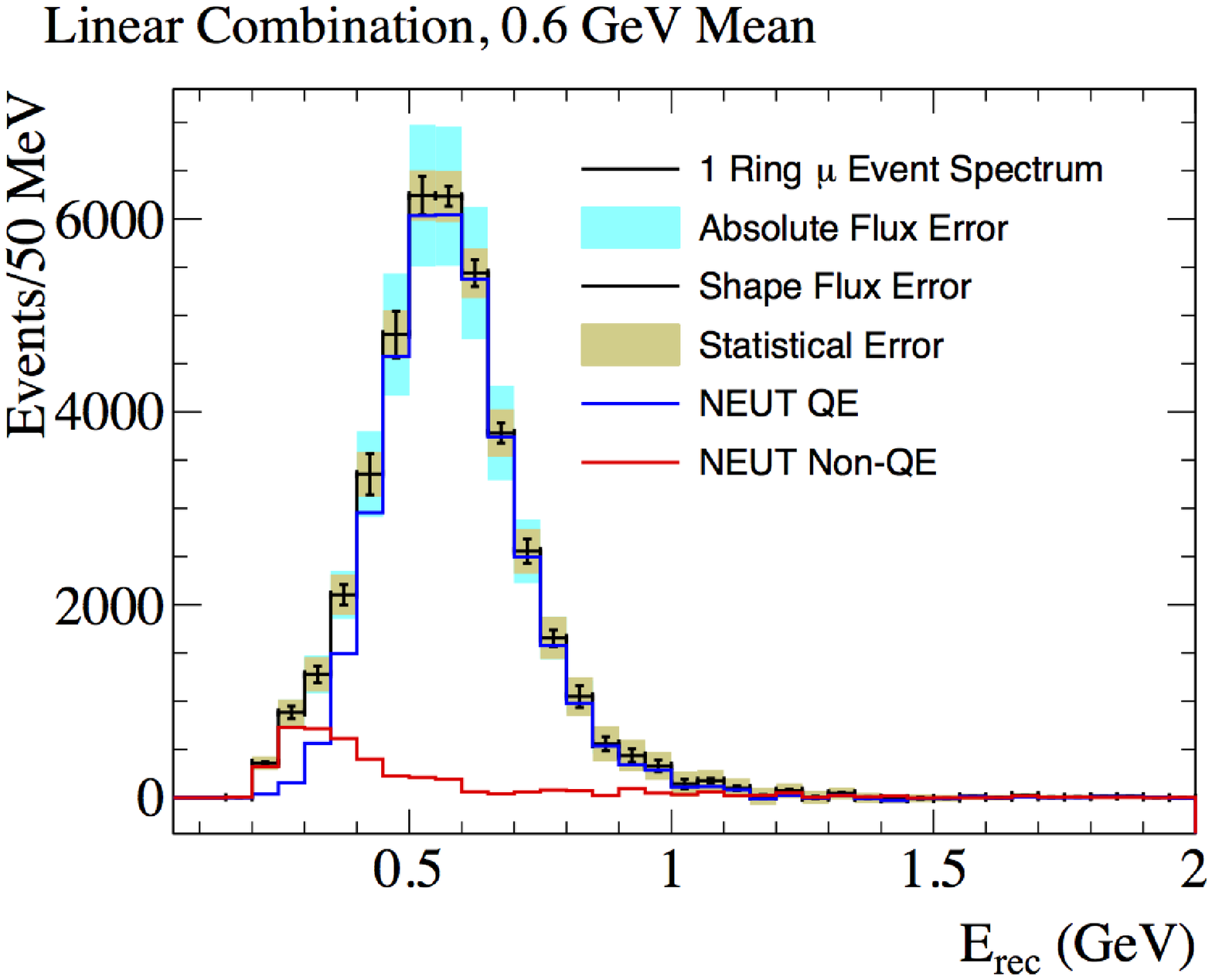}
\centering\\
a) 600~MeV.
\end{minipage}\hfill
\begin{minipage}{.46\textwidth}
\centering
\includegraphics[height=5cm]{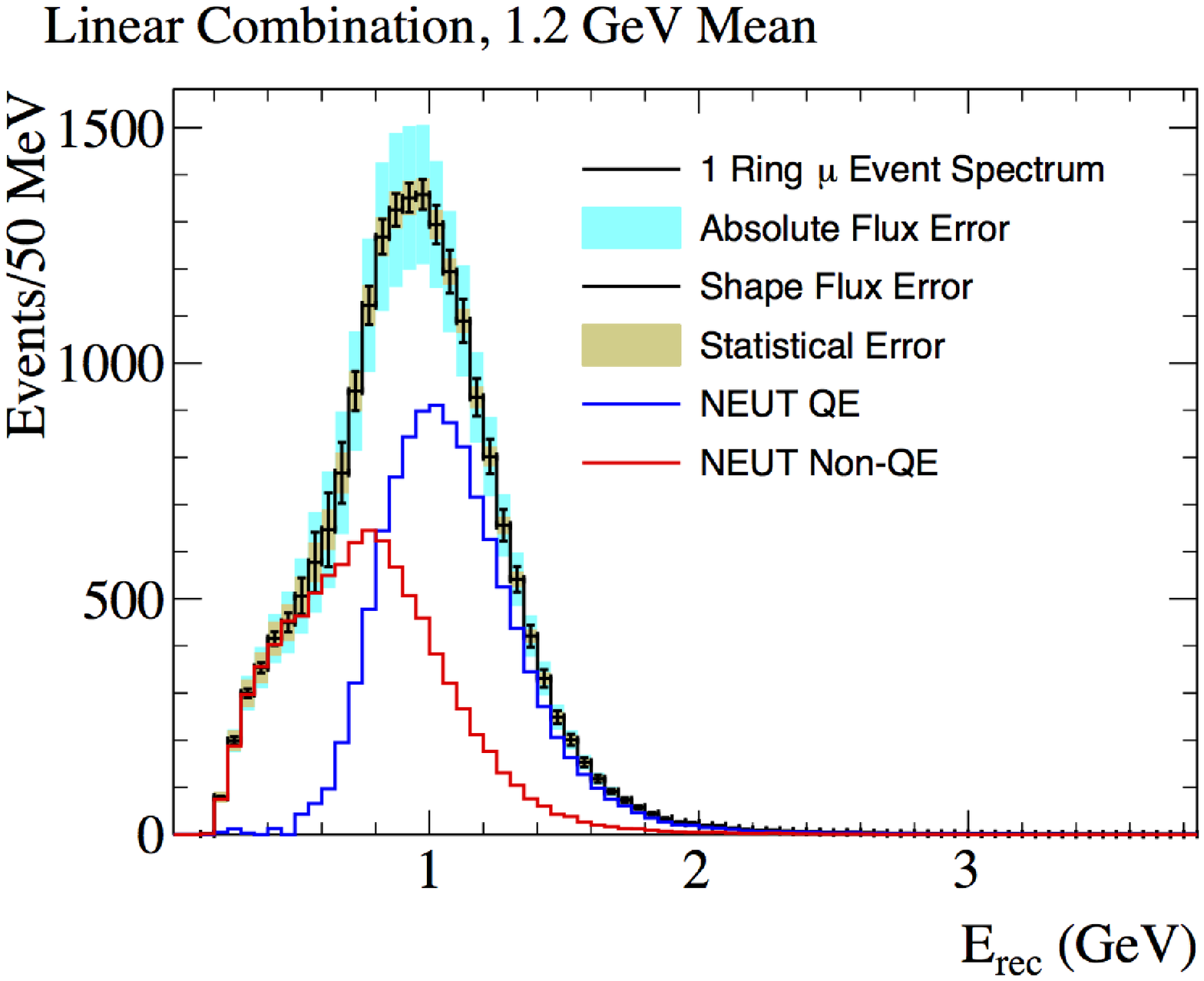}\\
b) 1200~MeV.
\end{minipage}
\caption[margin=5cm]{The reconstructed energy distributions for the events in Figure~\ref{fig:gaus_enu}.  The plots also include the expected distributions for true CCQE and non-CCQE events.}
\label{fig:gaus_erec}
\end{figure}

The same event samples are shown as a function of reconstructed neutrino energy in Figure~\ref{fig:gaus_erec}, where the reconstructed energy is calculated assuming the observed lepton was produced from a CCQE interaction on a single nucleon at rest.  These plots also include the expected distribution for all true CCQE and non-CCQE events in the MC, demonstrating clear separation between CCQE and non-CCQE in the selected event samples.

\begin{figure}
\begin{minipage}{.46\textwidth}
\includegraphics[height=5cm]{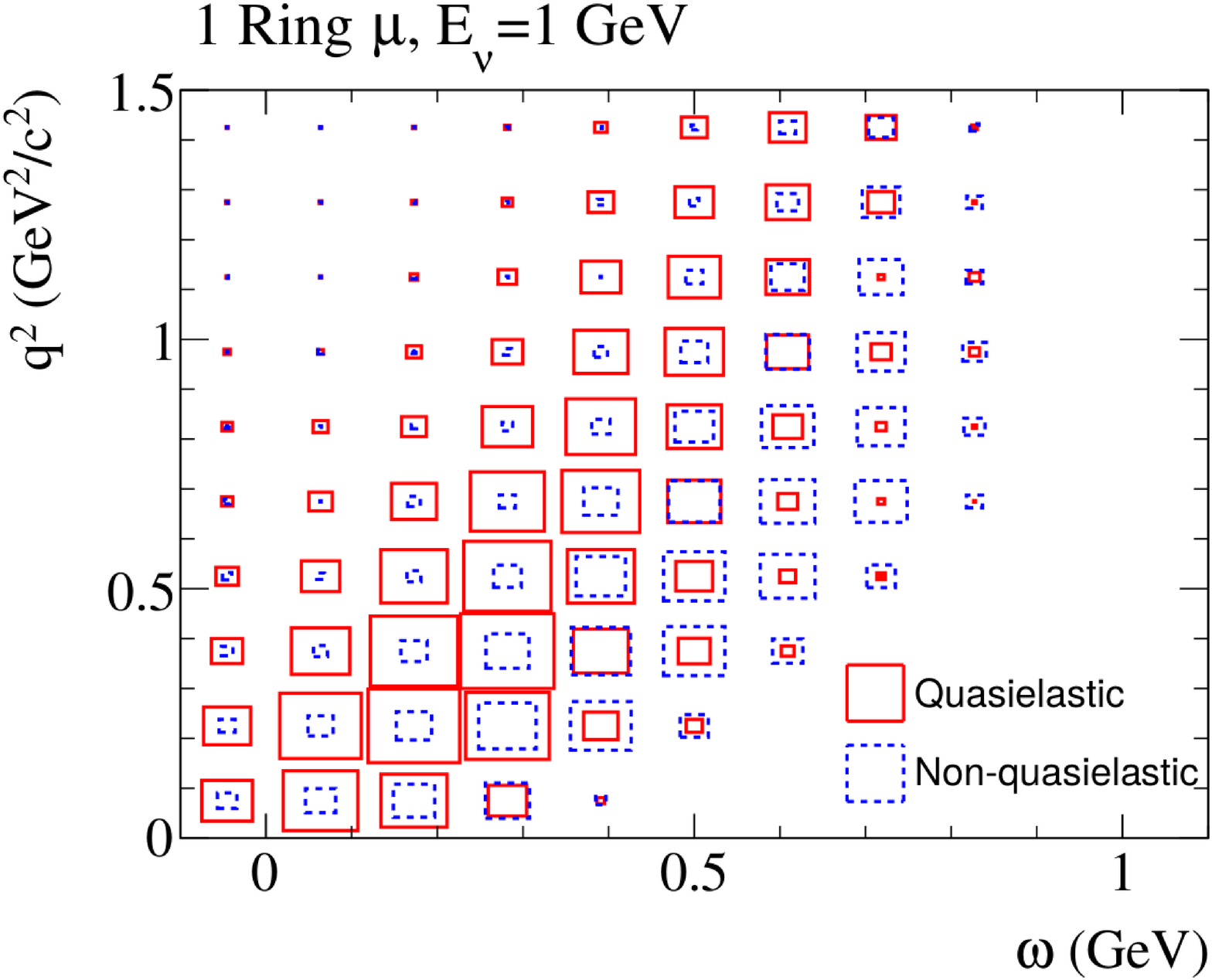}
\centering
\caption[margin=5cm]{The reconstructed 3-momentum transfer (q$^{2}$) plotted against the reconstructed energy transfer ($\omega$) for selected events from a Gaussian neutirno flux centred at 1~GeV.  True CCQE events are shown by the red boxes, with the dashed blue indicating non-CCQE events.}
\label{fig:q2_omega}
\end{minipage}\hfill
\begin{minipage}{.46\textwidth}
\centering
\includegraphics[height=5cm]{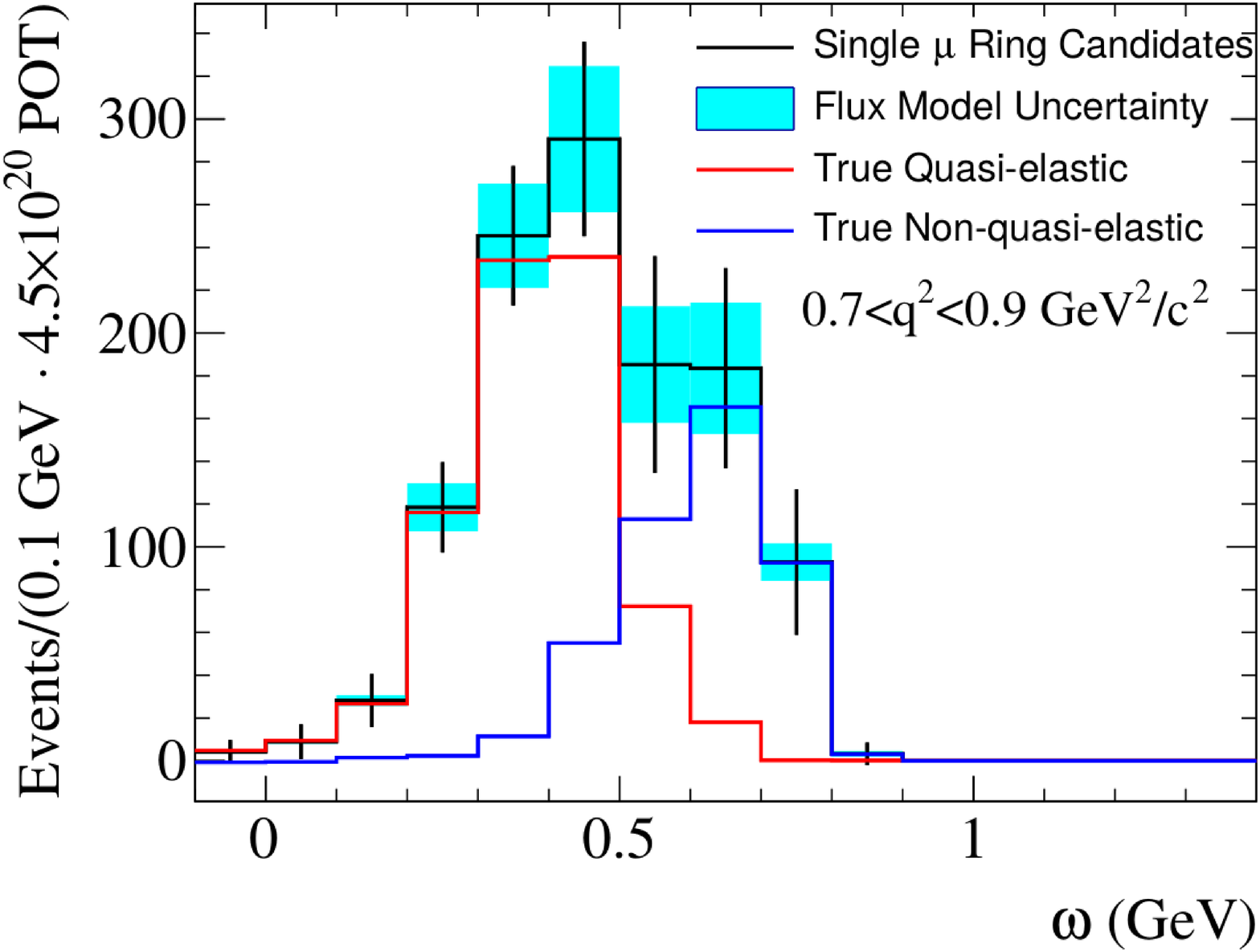}
\caption[margin=5cm]{A 1-D slice from Figure~\ref{fig:q2_omega} for q$^{2}$ values between 0.7 and 0.9~GeV$^{2}$/c$^{2}$, showing clear separation between CCQE (red) and non-CCQE (blue) events.}
\label{fig:gaus_omega}
\end{minipage}
\end{figure}

This measurement can be repeated using any reconstructed distribution of interest, such as neutron multiplicity, for Gaussian fluxes covering a range of neutrino energies.  These distributions would show how the quantity of interest changed with true neutrino energy, and would have highly correlated flux and detector uncertainties, something that would otherwise be impossible when averaging over the full neutrino flux energy spectrum.  Creating a known neutrino energy also allows analysers to perform neutrino scattering measurements as a function of the 3-momentum or energy transfer to the nucleus, shown in Figures~\ref{fig:q2_omega} and~\ref{fig:gaus_omega}.  These variables are better probes of neutrino scattering than neutrino energy, since they directly determine which interaction processes can take place.  This would mirror the methods used in electron scattering experiments, providing more accurate measurements of specific neutrino interaction processes.

\subsubsection{Short baseline oscillations}

\begin{figure}[ht!]
\centering
\begin{minipage}[t]{.46\textwidth}
\includegraphics[height=5cm]{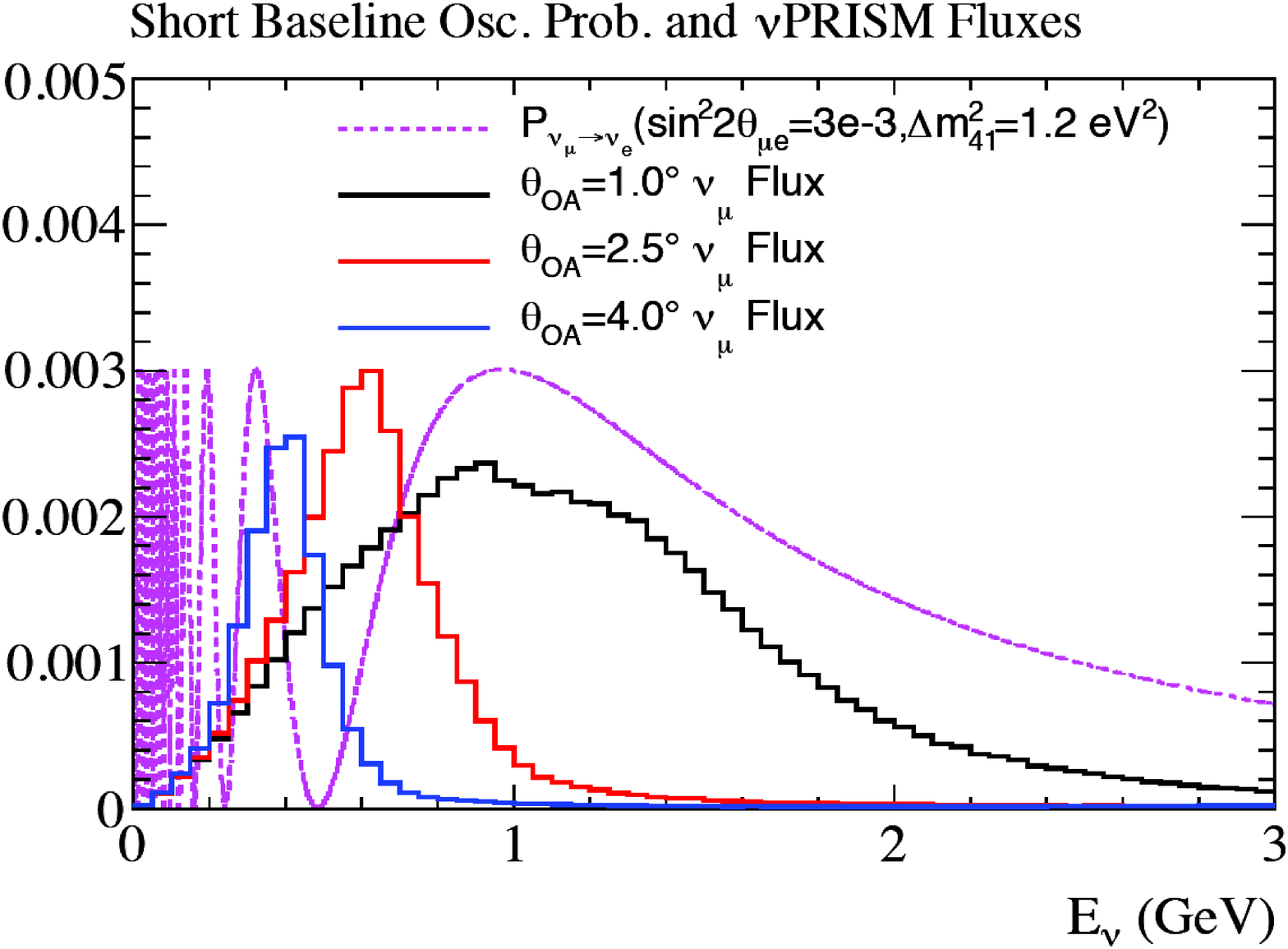}
\caption[margin=5cm]{The $\nu_{\textrm{e}}$ appearance probability at NuPRISM for the given values of the sterile oscillation parameters.  This is compared to the neutrino energy spectrum at three different off-axis angles, showing the change in oscillation probability across the NuPRISM tank.}
\label{fig:sterile_osc}
\end{minipage}
\end{figure}

Both TITUS and NuPRISM would have the correct baseline and neutrino energy spectrum to test the LSND~\cite{PhysRevD.64.112007} and MiniBooNE~\cite{Aguilar-Arevalo:2013pmq} short baseline results, but the NuPRISM concept provides some unique capabilities.  As Figure~\ref{fig:sterile_osc} shows, moving further off-axis in NuPRISM means that the neutrino spectrum being sampled peaks at different energies, which can be used to test the energy dependence of any oscillation signal.  The expected backgrounds also change with neutrino energy, but in a different way to the oscillated signal events.  

The NuPRISM analysis selects single ring, electron-like, events across all the off-axis slices and includes the full T2K flux and neutrino interaction uncertainties as described earlier.  These samples are fit as a function of reconstructed neutrino energy and off-axis angle to determine their sensitivity to the appearance of electron neutrinos from sterile oscillations.  More details are given in Ref.~\cite{Bhadra:2014oma}.

\begin{figure}
\begin{minipage}{.46\textwidth}
\includegraphics[height=5cm]{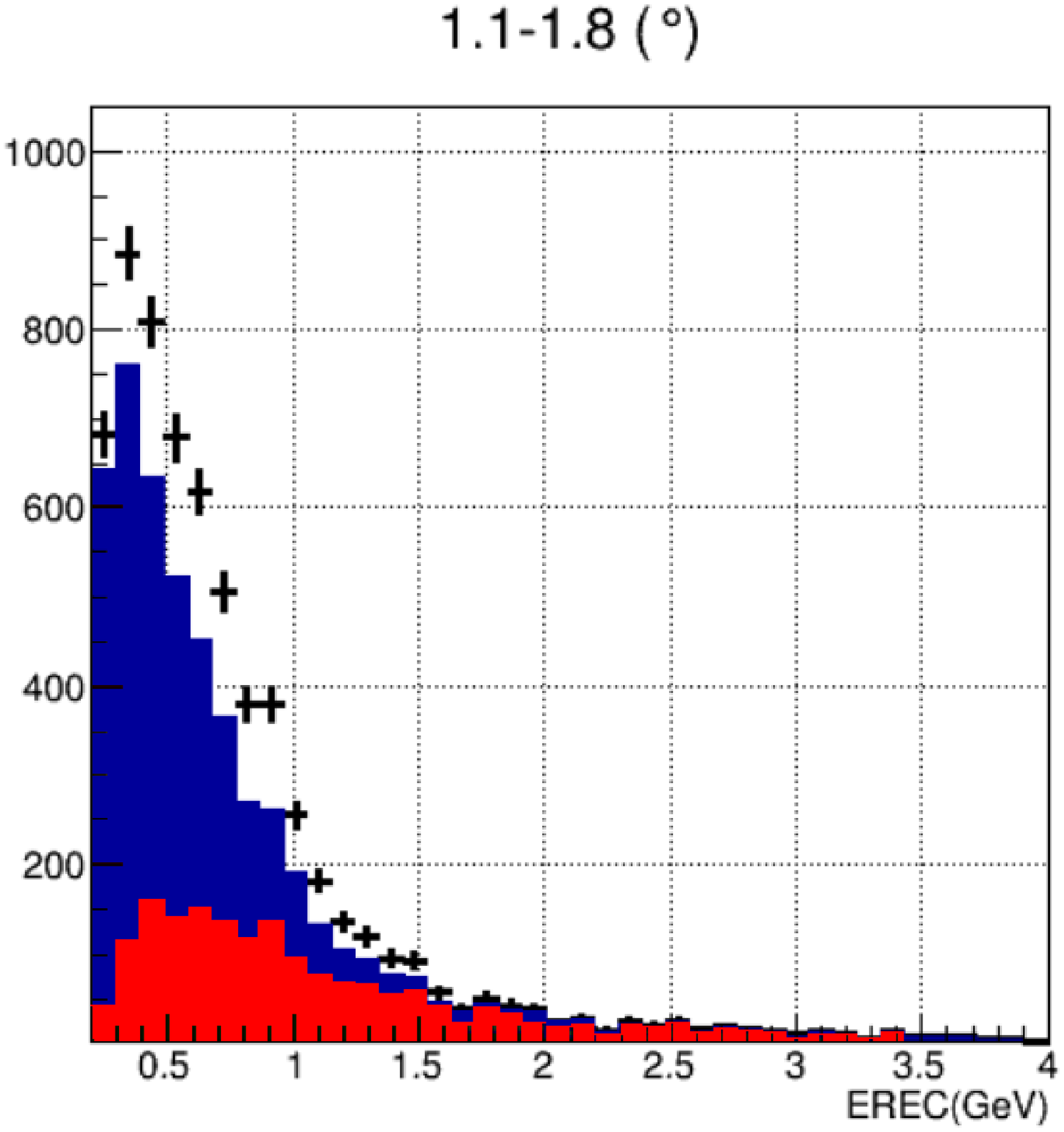}
\centering\\
a) The 1.1 -- 1.8~$^\circ$ on-axis slice.
\end{minipage}\hfill
\begin{minipage}{.46\textwidth}
\centering
\includegraphics[height=5cm]{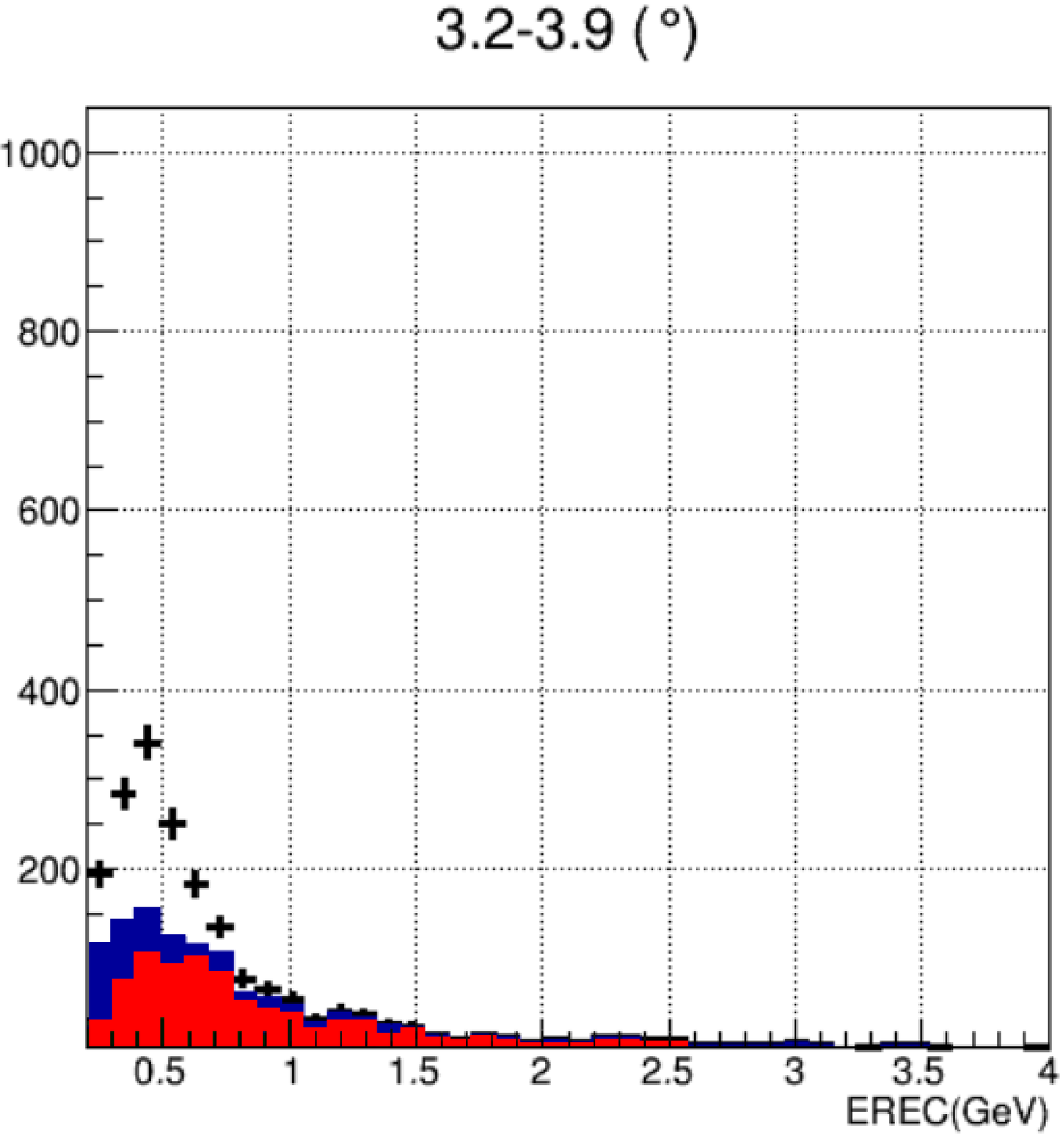}\\
b) The 3.2 -- 3.9~$^\circ$ off-axis slice.
\end{minipage}
\caption[margin=5cm]{The intrinsic electron neutrino background (red), muon neutrino background (blue) and appearance signal (points) for the NuPRISM sterile oscillation analysis searching for $\nu_{\textrm{e}}$ appearance.}
\label{fig:nuprism_bkg}
\end{figure}

The real power of NuPRISM is shown in Figure~\ref{fig:nuprism_bkg}, which plot the selected events in the most off-axis and most on-axis slices of the detector.  The blue histogram shows the selected background events coming from $\nu_{\mu}$ interactions, the red histogram shows the intrinsic beam $\nu_{\textrm{e}}$ background and the black points are a given appearance signal. In the on-axis slice there is a large $\nu_{\mu}$ contamination and the signal events have a broad reconstructed energy distribution.  Moving to the off-axis slice the $\nu_{\mu}$ contamination is greatly reduced whilst the signal is concentrated in a narrow reconstructed energy region.

\begin{figure}
\begin{minipage}{.46\textwidth}
\includegraphics[height=5cm]{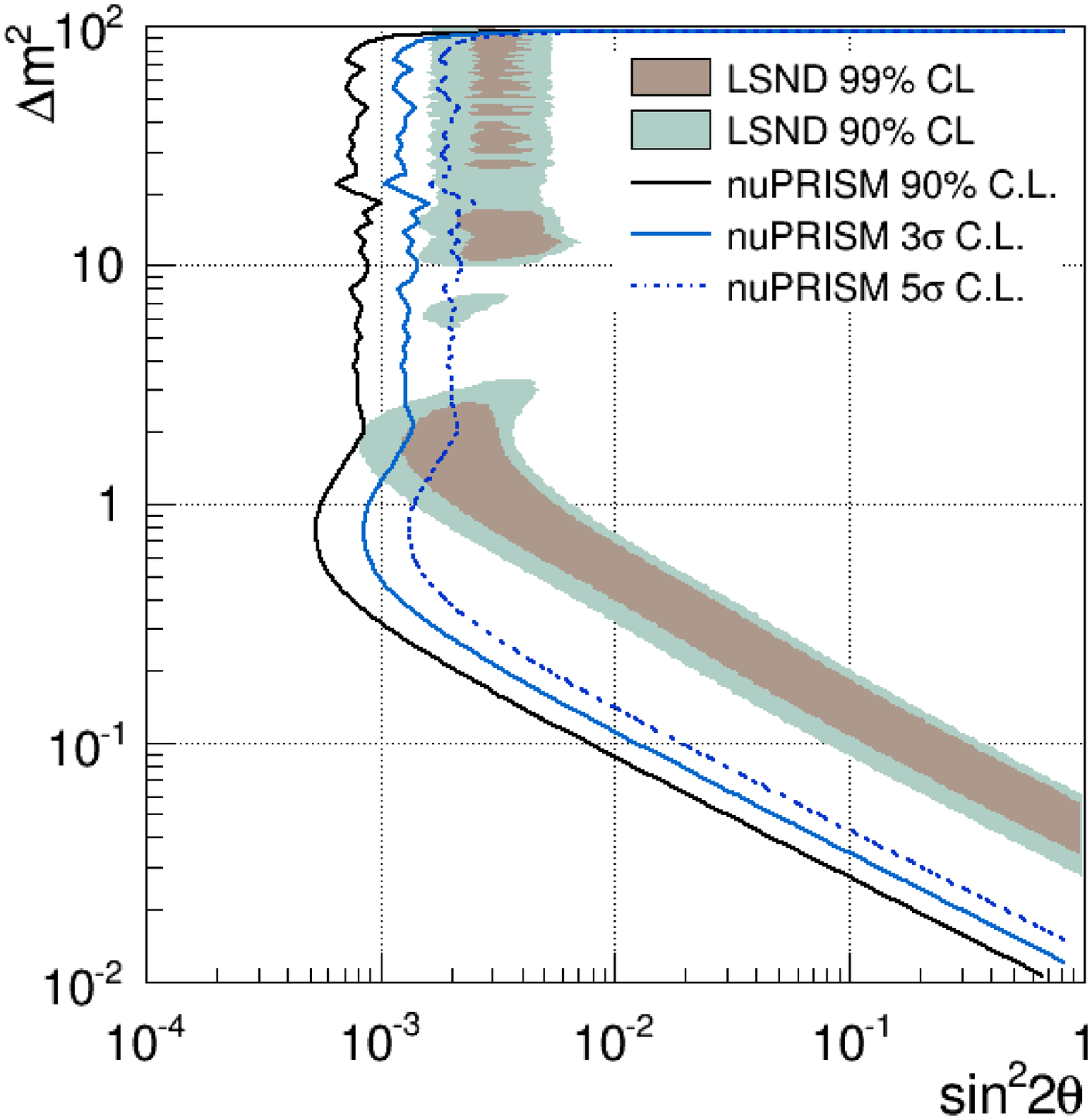}
\centering\\
a) Exclusion region with 2.25$^{21}$ POT in neutrino beam mode, the expected T2K exposure
\end{minipage}\hfill
\begin{minipage}{.46\textwidth}
\centering
\includegraphics[height=5cm]{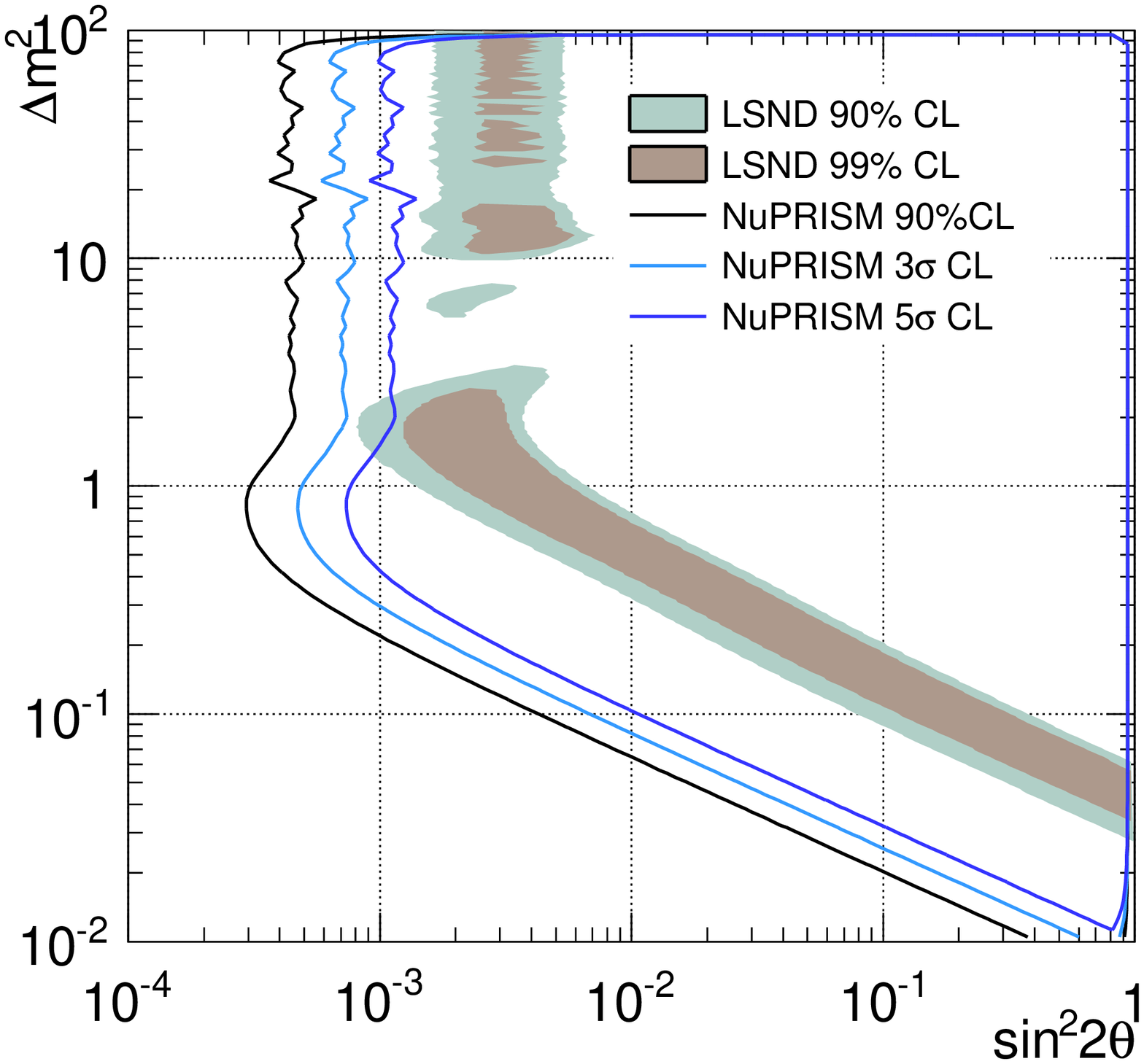}\\
b) Exclusion region with 7.5e$^{21}$ POT in neutrino beam mode, the expected T2K-II exposure
\end{minipage}
\caption[margin=5cm]{The oscillation parameter regions where NuPRISM can exclude sterile neutrino oscillations at the 90\%, 3~$\sigma$ and 5~$\sigma$ confidence level.  This is compared to the allowed region from the LSND experiment.}
\label{fig:sterile_sens}
\end{figure}

This behaviour allows NuPRISM to set strong constraints on the sterile oscillation parameter phase space, shown in the left plot of Figure~\ref{fig:sterile_sens}.  This shows NuPRISM excluding the entire LSND allowed region at 90\% confidence, with most of it excluded at $5~\sigma$.  This is expected to improve for future analyses, which will use a full detector simulation and reconstruction to provide increased statistics data samples and direct constraints on the background processes.  The effect of an increase in statistics is shown in the right plot of Figure~\ref{fig:sterile_sens}, where the analysis has been re-done assuming the T2K-II exposure, greatly increasing the excluded region across the parameter space.  This analysis will be further improved by the inclusion of the existing T2K near detector, allowing a full near-far oscillation analysis at a short baseline.

\section{Summary}

The current generation of long baseline neutrino oscillation experiments have reached the point where systematic uncertainties have a noticeable effect on their measurements.  To make a measurement of CP violation in the lepton sector requires a solid understanding and good control of these systematics, something not possible with the current T2K near detector.  Building an intermediate water Cherenkov detector will address the shortcomings of the ND280, reducing the uncertainty for the T2K-II and Hyper-K oscillation experiments.  The NuPRISM detector has a compelling physics program in addition to this, providing unique measurements of neutrino scattering and a powerful probe of short baseline neutrino oscillations.

%\bibliography{references}{}
%\bibliographystyle{unsrt}

\providecommand{\noopsort}[1]{}\providecommand{\singleletter}[1]{#1}%

\end{document}